\documentclass{elsart}
\usepackage{epsfig}
\usepackage{amsmath}
\usepackage{amssymb}
\usepackage{dcolumn}
\usepackage{graphicx}
\usepackage{dcolumn}

\begin{document}
\begin{frontmatter}

\title{Effective Particle Methods for Fisher-Kolmogorov Equations: Theory and Applications to Brain Tumor Dynamics}

\author{Juan Belmonte-Beitia}

\address{Departamento de Matem\'aticas, E. T. S. I. Industriales and Instituto de Matem\'atica Aplicada a la Ciencia y la Ingenier\'{\i}a (IMACI), 
Universidad de Castilla-La Mancha, 13071 Ciudad Real, Spain ({\tt juan.belmonte@uclm.es}). }

\author{Gabriel F. Calvo}
\address{Departamento de Matem\'aticas, E. T. S. I. Caminos, Canales y Puertos and Instituto de Matem\'atica Aplicada a la Ciencia y la Ingenier\'{\i}a (IMACI), Universidad de Castilla-La Mancha, 13071 Ciudad Real, Spain ({\tt gabriel.fernandez@uclm.es}).}

        \author{V\'{\i}ctor M. P\'erez-Garc\'{\i}a}
        \address{Departamento de Matem\'aticas, E. T. S. I. Industriales and Instituto de Matem\'atica Aplicada a la Ciencia y la Ingenier\'{\i}a (IMACI), Universidad de Castilla-La Mancha, 13071 Ciudad Real, Spain  ({\tt victor.perezgarcia@uclm.es}).}

\date{\today}

\begin{abstract}
Extended systems governed by partial differential equations can, under suitable conditions, be approximated by means of sets of ordinary differential equations for global quantities capturing the essential features of the systems dynamics. Here we obtain a small number of effective equations describing the dynamics of single-front and localized solutions of Fisher-Kolmogorov type equations. These solutions are parametrized by means of a minimal set of time-dependent quantities for which ordinary differential equations ruling their dynamics are found. A comparison of the finite dimensional equations and the dynamics of the full partial differential equation is made showing a very good quantitative agreement with the dynamics of the partial differential equation. We also discuss some implications of our findings for the understanding of the growth progression of certain types of primary brain tumors and discuss possible extensions of our results to related equations arising in different modelling scenarios.
\end{abstract}


\end{frontmatter}

\section{Introduction}

Many partial differential equations of relevance in applied sciences have robust localized solutions displaying particle-like behavior. A wealth of distinct behaviors are encountered and in general they are referred to as coherent structures and/or solitary waves\cite{General1,General2,General3,General4}.
\par
In a limited number of prominent cases the equations are known to be integrable. When that happens, there is an infinite number of conserved quantities and the solution of the initial value problem can be constructed using different mathematical methods such as the inverse scattering transform. In integrable systems initial data can be rigorously decomposed into solitons plus radiation (linear modes) and a complete analysis of the asymptotic dynamics can be made.
\par
While there are several physically relevant systems ruled by integrable partial differential equations, there is a vast majority of problems that are nonintegrable. Remarkably, some of them consist of small perturbations to integrable problems. In those cases one can still construct a rigorous theory for the dynamics of solitons that allows for an analytical description of the dynamics \cite{General4,Kivshar1}. However, in many other instances the perturbations are not ``small" and/or the basic underlying problem is not integrable but coherent structures still persist and constitute a basic elemente in the dynamics. 
\par
In many of those problems  a variational formulation can be written and then a very popular method is the, so-called, effective Lagrangian method, collective coordinate method or effective particle method. This method assumes the profile of the solution to be given by a specific ansatz depending on a small number of time dependent parameters. The specific choice of the ansatz depends on the equation under study and  in many cases is suggested by physical considerations.  The names ``effective particle" and ``collective coordinates" come from the fact that, in the framework of this approach, one simplifies the dynamics of an extended field with spatio-temporal dependencies, i.e. having ``infinite" degrees of freedom" to a finite (small) number of time-dependent quantities (coordinates). The name comes from analogy with classical mechanics that provides a simple description (coordinates) of a typically extended object. 

While the ansatz does not provide an exact solution of the PDE it is tipically used through the variational formulation as a test function and equations are obtained for the evolution of the solitary wave parameters. This approach works remarkably well for many relevant problems having Hamiltonian structure and provides a way  to describe the infinite-dimensional dynamics in a simple form when the dynamics is dominated by coherent structures. This method has been exploited extensively in applied sciences in a large number of works for a broad variety of equations having a variational formulation (see e.g. Refs. \cite{General1,Var1,Var2,Var3,Var4,Var5,Var6,Var7} and references therein). When coherent structures are robust it furnishes a simple description of the dynamics. The main weaknesses of the effective particle method are that: (i) it requires some experience to select appropriate ansatzes that capture adequately the dynamics, and  (ii) the reduction to finite dimensions is provided without a measure of the error of the approximation that is estimated a posteriori on the basis of numerical simulations of the parent PDE. While the method has been used in hundreds of papers dealing with the dynamics of nonlinear waves in non-integrable systems, to our knowledge there are no papers obtaining a priori error bounds for such types of approximation.

\par

Of particular interest are those equations that cannot be derived from a variational principle as it happens e.g. in dissipative systems. In that context there has been a great interest on different types of finite-dimensional descriptions of the dynamics of systems ruled by PDEs in a variety of contexts (see e.g. \cite{Red4,Red1,Red2,Red3} and references therein). However a simple procedure such as the one provided by effective particle methods that allows applied scientists to reduce the dynamics of a partial differential equation with solitary waves to a set of finite dimensional simple equations for the solitary wave parameters is not available yet. In this paper we present a very simple methodology that allows to obtain those types of approximations for  the Fisher-Kolmogorov (FK)
%
%
and related reaction-diffusion equations. The simplest version of the FK equation is
\begin{equation}\label{dimensionalFK}
u_{t}=Du_{xx}+ \rho u(1-u), 
\end{equation}
and describes the evolution of a population density $u(x,t)$ measured in units of a maximal population $u_*$ on a given spatial domain. This 
equation is the simplest reaction diffusion model incorporating two effects: dispersion with a dispersal rate $D>0$ and proliferation or population growth with rate $\rho>0$. In Eq. (\ref{dimensionalFK}) the population growth $g(u) = \rho u (1-u)$ is of the so-called logistic type although other terms $g(u)$ with similar qualitative form have been used in the literature. Eq. (\ref{dimensionalFK}) is written here in dimensional form in order to connect better with applications, although one can rescale the spatial and temporal variables to get rid of the coefficients $D$ and $\rho$.

\par

The FK equation and its extensions are a family of ubiquous reaction-diffusion models  arising  in population dynamics problems \cite{Murray,Shigesada,PP}, most prominently in cancer modelling \cite{Swanson1,Swanson2,PG1}, in the description of propagating crystallization/polymerization fronts \cite{Genzer2007}, chemical kinetics \cite{Gen3}, geochemistry \cite{Gen4} and many other fields (see e.g. \cite{Gen1,Gen2} and references therein). These equations do not admit a Lagrangian density depending on the field $u$ \cite{General2} and thus the variational formulation for the effective particle parameters cannot be written in the usual way.
\par

In this paper we get a small set of ordinary differential equations mimicking, not only the asymptotic dynamics of fronts arising in the Fisher-Kolmogorov (FK) equation but also describing their transient evolution towards the asymptotic regime. The plan of the paper is as follows: First, in section \ref{method}, we present the theoretical approach and discuss its application to the find the evolution of kink-like initial data classical FK equation. A comparison of  the results of the effective particle method with the numerical solution of the FK equation is made. Next, in section \ref{localized_method}, we apply the method to get  effective equations for the dynamics of initially localized solutions to the FK equation. We identify different dynamical regimes for three relevant quantities associated to the spatio-temporal evolution of such localized profiles of the FK equation which are not apparent from the usual numerical solution. Secs. \ref{sec-apl} and \ref{sec-apl2} are devoted to several applications of the method relevant for the understanding of the growth dynamics of certain types of  brain tumors. Next, in Sec. \ref{sec-V} we present an example of applications to models beyond the FK equation by adding a spatial dependence to the diffusion coefficient. Finally, in section \ref{discussion}, we discuss the implications of our results and summarize the conclusions.

\section{Effective-particle method description of front-type solutions}
\label{method}

\subsection{Derivation of effective equations for the front parameters}

Nonnegative single-front-type travelling wave solutions $u=u(z=x-ct)$ of Eq. \eqref{dimensionalFK} satisfying the boundary conditions
\begin{equation}\label{bc}
\lim_{z\rightarrow-\infty}u(z)=1,\quad \lim_{z\rightarrow+\infty}u(z)=0,
\end{equation}
obey the ODE
\begin{equation}\label{TW}
Du''(z)+cu'(z)+\rho u(1-u)=0 ,
\end{equation}
and have been studied in detail \cite{Murray}. It is well known that such fronts can be constructed whenever $c \geq 2\sqrt{\rho D}$ and a celebrated result by Kolmogorov and coworkers \cite{KKK} states that compact support initial data decay asymptotically into this type of waves with $c_\textrm{min}\equiv 2\sqrt{\rho D}$. However, less is known on the transient dynamics until the asymptotic regime is reached (see e.g. \cite{Sherrat} and references therein).  

\par

Following the basic idea behind effective particle methods, our aim is to find an approximation for the dynamics of Eq. \eqref{TW} by means of a simple finite-dimensional expression of the form
\begin{equation}\label{fp}
u(x,t)=A(t)f\!\left(\frac{x-X(t)}{w(t)}\right) .
\end{equation}

The ``effective particle" describing the front is parametrized by three quantities depending only on time, namely, the wave amplitude $A=A(t)$, the front position $X=X(t)$ and the width $w(t)$.

A key point of the method is the choice of a suitable profile function $f$ in Eq. \eqref{fp} approximating the spatial profile of the solution. In our case, we will take the profile in \eqref{fp} to be inspired by the Ablowitz solution \cite{Ablowitz}
\begin{equation}\label{solFK}
u(z)=\frac{1}{(1+e^{z/\sqrt{6}})^2}.
\end{equation}
The solution given by Eq. (\ref{solFK}) is the only simple explicit solution known for the Fisher-Kolmogorov equation (in its adimensional version, i.e. with $\rho = D=1$), but corresponds to the specific speed $c=5/\sqrt{6}$, slightly larger than the minimal speed solution to the FK equation.
 Thus,
a natural choice for our front profile is 
\begin{equation}\label{profile}
u(x,t)=\frac{A(t)}{\left[1+e^{(x-X(t))/w(t)}\right]^2} \, ,
\end{equation}
that has the expected asymptotic exponential decay for large values of $x$. In what follows we will try to obtain equations for the dynamics of the ``effective particle" defined by the parameters $A(t), w(t), X(t)$.

\par

To proceed with the method, let us define the integral quantities:
\begin{subequations}
\label{Is}
\begin{eqnarray}
I_{1}(t)&=&\int_{-\infty}^{\infty}u_{x}dx,\label{norma}\\
I_{2}(t)&=&\frac{\int_{-\infty}^{\infty}xu_{x}dx}{I_{1}(t)},\label{cm}\\
I_{3}(t)&=&\frac{\int_{-\infty}^{\infty}(x-I_{2}(t))^{2}u_{x}dx}{I_{1}(t)},\label{anchura}
\end{eqnarray}
\end{subequations}
which are related to the $L_{1}$-norm (number of particles), center of mass and width of the gradient of the density $u$, respectively. 

Then, introducing \eqref{profile} in integral \eqref{norma}, it follows that $I_{1}(t)=-A(t)$. The evolution of $I_1(t)$ can be obtained by a direct formal calculation
\begin{multline}
\frac{dI_{1}}{dt} =  \frac{d}{dt}\left( \int_{-\infty}^{\infty} u_{x}dx \right)= \int_{-\infty}^{\infty} u_{xt}dx = \int_{-\infty}^{\infty}\left(Du_{xxx}+\rho u_{x}-2\rho uu_{x}\right)dx  \\
  = -\rho A(t)+4\rho A^{2}(t)\int_{-\infty}^{\infty}\frac{e^{z}}{(1+e^{z})^5}dz=-\rho A(t)\left[1-A(t)\right],
\end{multline}
where we have used Eq. \eqref{dimensionalFK} and the fact that $\displaystyle{\int_{-\infty}^{\infty}e^{z}/(1+e^{z})^5dz=1/4}$.
\par
 We may calculate $I_{2}(t)$ in a similar way to get
\begin{equation}\label{I2}
I_{2}(t)=X(t)-w(t) .
\end{equation}
On the other hand, differentiating $I_{2}(t)$ with respect to time and using Eq. \eqref{dimensionalFK}, we find
\begin{eqnarray}
\frac{dI_{2}}{dt}&=&  -\frac{I_{2}(t)}{I_{1}(t)}\frac{dI_{1}}{dt} + \frac{1}{I_{1}(t)}\int_{-\infty}^{\infty} x\left(Du_{xxx}+\rho u_{x}-2\rho uu_{x}\right)dx = \frac{5}{6}\rho A(t)w(t),
\label{dI2dt}
\end{eqnarray}
where we have taken into account that $\displaystyle{\int_{-\infty}^{\infty}ze^{z}/(1+e^{z})^5dz=-11/24}$. Thus, differentiating \eqref{I2} and combining it with \eqref{dI2dt}, it follows that
\begin{equation}\label{evcm}
\frac{dX}{dt} = \frac{dw}{dt} + \frac{5}{6}\rho A(t)w(t).
\end{equation}
To get an equation for the time evolution of the width $w(t)$ we can first use Eq. (\ref{anchura}) to obtain
\begin{equation}\label{I3}
I_{3}(t)=\left( \frac{\pi^2}{3} - 1\right)\!w^{2}(t).
\end{equation}
\par
The time evolution of the width can be derived once more from the FK equation
\begin{multline}\label{dI3dt}
\frac{dI_{3}}{dt} = -\frac{I_{3}(t)}{I_{1}(t)}\frac{dI_{1}}{dt} + \frac{1}{I_{1}(t)}\int_{-\infty}^{\infty} \left[ x- I_{2}(t)\right]^{2}\left(Du_{xxx}+\rho u_{x}-2\rho uu_{x}\right)dx \\ 
 = 2D - \frac{1}{3}\rho A(t)w^{2}(t) .
\end{multline}
\par
Therefore, using expression \eqref{I3} for $I_{3}(t)$ together with Eq. \eqref{dI3dt}, the time evolution of the width $w(t)$ easily follows. 
\par
Summarizing, we get the following set of differential equations for the evolution of the front parameters as described by Eq. (\ref{fp})
\begin{subequations}
\label{ODEs}
\begin{eqnarray}
\frac{dA}{dt} & = & \rho A(t)\left[1-A(t)\right]\! , \label{Ampl} \\
\frac{dX}{dt} - \frac{dw}{dt}& = & \frac{5}{6}\rho A(t)w(t), \label{Veloc}\\
\frac{dw^{2}}{dt} & = & \frac{6D}{\pi^{2}-3}-\frac{\rho}{\pi^{2}-3}A(t)w^{2}(t). \label{Width}
\end{eqnarray}
\end{subequations}

Thus, our method provides a simple set of differential equations governing the evolution of a few (but very noteworthy) quantities describing the propagation of the front. As it will be described in Sec. \ref{analy}, explicit solutions for Eqs. (\ref{ODEs}) can be found what means that the dynamics of the reduced system can be easily found in closed form. 
Since Eq. (\ref{dimensionalFK}) is time invariant and consequently Eqs. (\ref{ODEs}) autonomous we will choose without loss of generality $t_0 = 0$ in what follows without loss of generality. Since ``a priori" error estimates are not available for our approach, we will later verify the quality of the approximation by resorting to numerical simulations to compare the results obtained from the PDE \eqref{dimensionalFK} with the reduced ODE model of Eqs. \eqref{ODEs}. 

\subsection{Analytical solutions}
\label{analy}

Although Eqs. (\ref{ODEs}) are a set of coupled nonlinear evolution equations their exact solutions can be found in closed form. First, Eq. (\ref{Ampl}) is a logistic equation whose solution is given by
\begin{equation}\label{Adet}
A(t) = \frac{A_0 e^{\rho t}}{1 + A_0 \left( e^{\rho t}-1\right)},
\end{equation}
where $A_0 = A(0)$. 
The expression for $A(t)$ given by Eq. (\ref{Adet}) can be inserted into Eq. \eqref{Width}, which is linear in $w^{2}(t)$, to obtain the explicit solution for $w(t)$ which reads (with $w(0) = w_0$)

\begin{eqnarray}
w^{2}(t)&=&\frac{w_0^{2} - \frac{6D}{\rho A_0}g(0)}{\left( 1 - A_0 + A_0e^{\rho t}\right)^{\frac{1}{\pi^{2}-3}}} + \frac{6D}{\rho A_0} \left( 1 - A_0+ A_0e^{\rho t}\right) e^{-\rho t} g(t)\, ,
\label{Solwidth}
\end{eqnarray}
where
\begin{eqnarray}\label{Solg}
g(t) = \sum_{n=0}^{\infty} \frac{\Gamma(n+1)\Gamma\left(1-\frac{1}{\pi^{2}-3}\right)}{\Gamma\left(n+1-\frac{1}{\pi^{2}-3}\right)} \left( 1 -\frac{1}{A_0}\right)^{n}e^{-n\rho t} \, ,
\end{eqnarray}
is the Gaussian hypergeometric function $F(\alpha,\beta;\gamma;\eta)$ with $\alpha=\beta=1$, $\gamma=1-\frac{1}{\pi^{2}-3}$, $\eta = \left( 1 -\frac{1}{A_0}\right)e^{-\rho t}$ and $\Gamma(z)=\int_{0}^{\infty} x^{z-1}e^{-x}dx$ the Euler gamma function. Notice that $g(t)\to1$ for $t\to\infty$. 
\par
Finally, we can also obtain the analytical expression for the velocity $v(t)$ by combining \eqref{Veloc} and \eqref{Solwidth} to get 
\begin{eqnarray}
v(t) = \frac{3D}{(\pi^{2}-3)w(t)}  + \left[\frac{5}{6} -\frac{1}{2(\pi^{2}-3)}\right] \frac{\rho A_0w(t)}{A_0+(1-A_0)e^{-\rho t}} .
\label{Solveloc}
\end{eqnarray}
Thus Eqs. (\ref{Adet}-\ref{Solveloc}) provide the full dynamics of the front for all times. We can easily find the  asymptotic behavior of these solutions to be
\begin{subequations}
\label{asymptoticparameters}
\begin{eqnarray}
A(t) & \underset{t \to \infty}{\longrightarrow} & 1,  \\
 w(t) & \underset{t \to \infty}{\longrightarrow} & \sqrt{\frac{6D}{\rho}} \simeq 2.45 \sqrt{\frac{D}{\rho}}, \\
v(t) & \underset{t \to \infty}{\longrightarrow} & 5\sqrt{\frac{\rho D}{6}} \simeq 2.04\sqrt{\rho D}.
 \end{eqnarray}
 \end{subequations}
Also, it is worth mentioning that despite the crudeness of the approximation of the effective particle method, i.e. that the front maintains its basic shape during the evolution, the velocity $v(t)$ in the asymptotic limit $t\to\infty$ is very close to the exact one $c_\textrm{min}=2\sqrt{\rho D}$, with the percentual relative error being of the order of 2\%, that is the difference between the real asymptotic speed and the one of the Ablowitz solution. However, the main strength of the method is that it provides qualitative information on the transient evolution of the front parameters.

\subsection{Comparison with the FK equation}

Due to the approximate nature of Eqs. (\ref{ODEs}) and the lack of a priori error bounds, it is necessary to validate 
the predictions of the effective particle description through a direct comparison with the numerical solution to Eq. \eqref{dimensionalFK}. To do so, we have solved numerically the Eq.  \eqref{dimensionalFK} with initial data given by \eqref{profile} and a particular set of 
 initial parameters $A_0, w_0, X_0$. From the solution $u(x,t)$ and using Eqs. \eqref{Is}, we obtain the soliton parameters numerically in terms of integral quantities, i.e. 
 \begin{subequations}
 \label{paramsPDE}
 \begin{eqnarray}
 A_{\textrm{PDE}}(t) & = & -I_{1}(t), \\
 w_\textrm{PDE}(t) & = & \sqrt{\frac{3I_{3}(t)}{\pi^{2}-3}},\\
 X_{\textrm{PDE}}(t) &= & w_{\textrm{PDE}}(t) + I_{2}(t).
 \end{eqnarray}
 \end{subequations}
 These values are to be compared with the solutions of Eqs. (\ref{ODEs}), i.e. with Eqs. (\ref{Adet}-\ref{Solveloc}).
 
\begin{figure}
 \begin{center}
\includegraphics[scale=0.42]{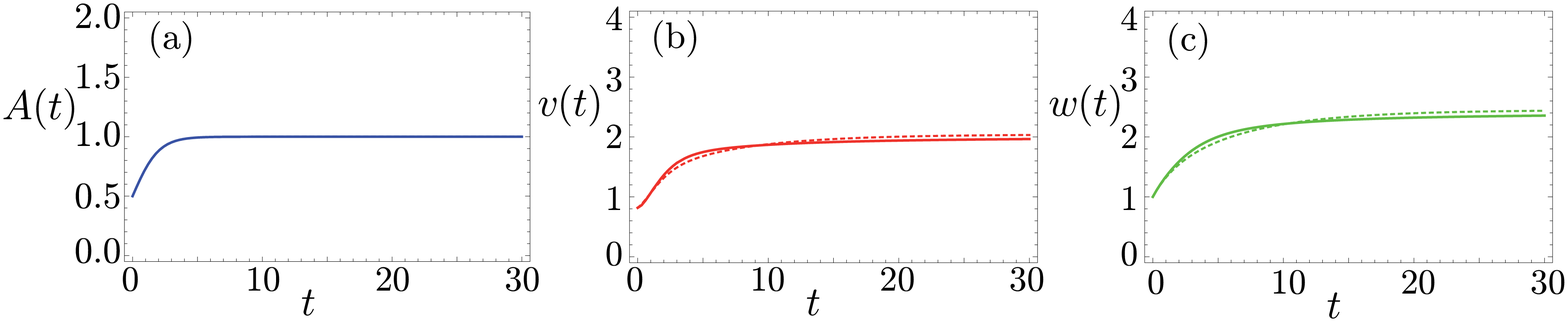}
\caption{[Color Online]. Comparison of the evolution of front solutions of the Fisher-Kolmogorov equation described by Eq. (\ref{dimensionalFK}) with 
the analytical solutions of the effective particle method given by Eqs. (\ref{Adet}-\ref{Solveloc}). The FK equation is solved numerically using a standard second order in space and time finite difference method with zero derivative boundary conditions and constants $D=1$ and $\rho=1$. The initial data is given by Eq. (\ref{profile}) with  $A_0 = 0.5$, $w_0 =1$, and $X_0 =2$. In all subplots (a)-(c) the solid lines correspond to the parameters \eqref{paramsPDE} extracted from the numerical solution of the FK equation. The dashed lines correspond to the analytical solutions of the ODEs. The subplots show the: (a) amplitude $A(t)$ (dashed) versus $A_{\textrm{PDE}}(t)$ (solid), (b) velocity of the front $v(t)$ (dashed) versus $v_{\text{PDE}}(t)$ (solid), (c) width of the solution $w(t)$ (dashed) versus $w_{\textrm{PDE}}(t)$ (solid).
\label{figura1}}
\end{center}
\end{figure}

Figure \ref{figura1} displays typical results of the comparative evolution of fronts according to our reduced model \eqref{ODEs} and from the Eq. (\ref{dimensionalFK}) through \eqref{paramsPDE}. The agreement between the reduced ODE set and the full PDE is excellent. To exclude the possibility of this result being the consequence of a fortunate choice of the initial data, we have explored different sets  of initial conditions in the range $0<A_0<1$, $0.5 \leq w_0\leq 5$, and $0.5 \leq X_0\leq 5$ and model parameters in the range $0.1 \leq D \leq 10$ and $0.1\leq \rho\leq 10$. In all cases a very good quantitative agreement among the three sets of curves is observed for all times, the percent relative errors being always smaller than $10\%$. The small (and expected) discrepancy of the asymptotic values for the speed and the width is always present in our calculations and is a result of our ansatz choice.
 
\par

\section{Effective particle methods for localized solutions}
\label{localized_method}

\subsection{Motivation}

While front solutions of the FK equation have relevance in many practical scenarios, there are cases where the solutions are initially localized. A typical example are models related to the propagation of tumors, that start from the onset as localized low amplitude cell densities and extend through the healthy tissue as localized solutions.  

To derive finite-dimensional simple models able to tackle these questions we may extend the effective particle method to obtain approximate localized solutions of the Fisher-Kolmogorov equation. The procedure, however, is less straightforward. The first aspect to be addressed is the choice of a proper ansatz.

\subsection{The choice of the ansatz}
In contrast with the profile given by Eq. \eqref{profile}, we now look for a nonnegative {\em two-front wave} $u=u(x,t)$ satisfying the boundary conditions 
\begin{equation}\label{bclocalized}
\lim_{x\rightarrow\pm\infty}u(x,t)=0,\quad \forall t>0 .
\end{equation}

Our simple finite-dimensional approximation will be of the form
\begin{equation}\label{fplocalized}
u(x,t)=A(t)\!\left[ f\!\left(\frac{x-X(t)}{w(t)}\right) - f\!\left(\frac{x+X(t)}{w(t)}\right) \right]^{2}\! ,
\end{equation}
with $f$ representing a single front. As before, the two counter-propagating fronts are parameterized by three quantities; the amplitude $A=A(t)$, the right-front position $X=X(t)$ and the front widths $w=w(t)$. We will further assume that the resulting profile is spatially symmetric $u(-x,t)=u(x,t)$ although this restriction can be lifted in systems without spatial symmetries. To this end, we resort to an extension of our previous ansatz
\begin{equation}\label{profilelocalized}
u(x,t)= A(t)\!\left[\frac{1}{1+e^{(x-X(t))/w(t)}} - \frac{1}{1+e^{(x+X(t))/w(t)}}\right]^2 .
\end{equation}

The above ansatz \eqref{profilelocalized} satisfies the following properties: first, if $X(t)=0$, then $u(x,t)=0$; next 
$u(-x,t)=u(x,t)$, for $t>0$; also $\lim_{x\rightarrow\pm\infty}u(x,t)=0$, for all $t>0$. Finally the amplitude at $x=0$ is given by
 $u(0,t)=A(t)\tanh^{2}\left[X(t)/2w(t)\right]$, for all $t>0$.
 
\par
\subsection{Evolution equations for the parameters of the effective particle}
 We now proceed to define the integral quantities:
\begin{subequations}
\begin{eqnarray}
n(t)&=&\int_{-\infty}^{\infty}u\, dx,\label{number}\\
\sigma^{2}(t)&=&\frac{1}{n(t)}\int_{-\infty}^{\infty}x^{2}u\, dx,\label{variance}\\
\gamma(t)&=&-\int_{0}^{\infty}u_{x}dx,\label{gamma}
\end{eqnarray}
\end{subequations}
These integral quantities are different from the ones ($I_1,I_2,I_3$) used to characterize the front solutions due to the fact that now we are dealing with localized solutions. The parameter $n(t)$
 represents the total ``mass"  and in population dynamics applications represents the normalized number of individuals. $\sigma^{2}(t)$ gives the variance of the density distribution having the biological meaning of spatial width or spatial extension occupied by the population. Finally, the parameter $\gamma(t)$ provides the right-front size, giving an estimate of the size of the infiltration region in applications.
\par
Upon substitution of Eq. \eqref{profilelocalized} in Eqs. \eqref{number}-\eqref{gamma}, and after integration, we get
\begin{subequations}
\begin{eqnarray}
n(t) &=& 2A(t)\!\left[ X(t)\coth\!\left(\frac{X(t)}{w(t)}\right) - w(t) \right],\label{numberint}\\
\sigma^{2}(t)&=& \frac{1}{3}X^{2}(t) + \frac{\pi^{2}}{3}w^{2}(t) - \frac{2X^{2}(t)w(t)}{3\!\left[ X(t)\coth\!\left(\frac{X(t)}{w(t)}\right) - w(t)\right]}\, ,\label{varianceint}\\
\gamma(t)&=&A(t)\tanh^{2}\!\left(\frac{X(t)}{2w(t)}\right).\label{gammaint}
\end{eqnarray}
\end{subequations}
\par
\begin{figure}
 \begin{center}
\includegraphics[scale=0.38]{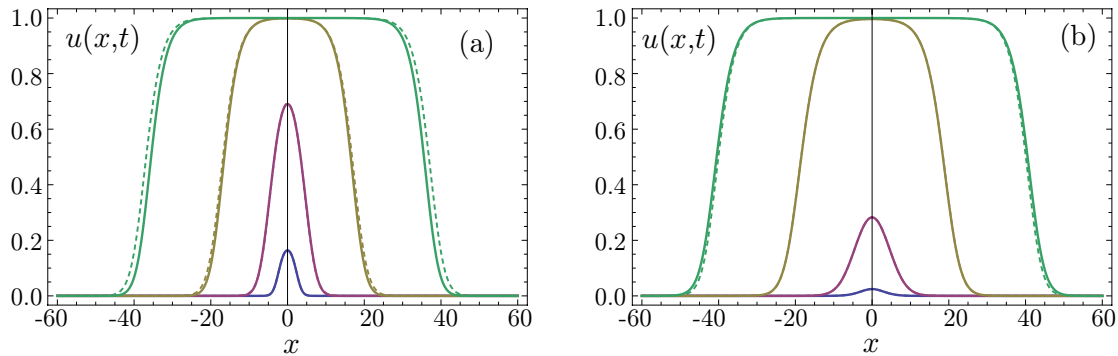}
\caption{[Color Online]. Comparison of the evolution of front solutions of the Fisher-Kolmogorov equation described by Eq. (\ref{dimensionalFK}) (solid curves) with 
that provided by the ansatz \eqref{profilelocalized} (dasher curves) with parameters given by  Eqs. \eqref{EDOn}-\eqref{EDOgamma} 
for various times: $t=0, 3, 10, 20$ from the innermost to the outermost profiles. The diffusion coefficient is $D=1$ and the growth rate $\rho=1$. The initial data is given by Eq. (\ref{profilelocalized}) with  (a) $A_0 = 0.2$, $w_0 =1$, and $X_0 =3$; (b) $A_0 = 0.9$, $w_0 =3$, and $X_0 =1$.
\label{figura2}}
\end{center}
\end{figure}
\par
The evolution of $n(t)$, $\sigma^{2}(t)$ and $\gamma(t)$ is obtained via the FK equation \eqref{dimensionalFK} as in the case of single-fronts. For $n(t)$, we find
\begin{eqnarray}
\hspace*{-3mm}
\frac{dn}{dt} &=& \frac{d}{dt}\left(\int_{-\infty}^{\infty} u\, dx \right) = \int_{-\infty}^{\infty} u_{t} dx = \int_{-\infty}^{\infty} \left[ Du_{xx}+ \rho u(1-u) \right]dx \nonumber\\
&=& \rho n(t) - \rho\int_{-\infty}^{\infty} u^{2}dx = 2\rho A(t)\!\left[ X(t)\coth\!\left(\frac{X(t)}{w(t)}\right) - w(t) \right] \nonumber\\
&-& \rho A^{2}(t)\!\left[ X(t)\coth\!\left(\frac{X(t)}{w(t)}\right)\!\!\left[ 2 + 5\textrm{csch}^{2}\!\left(\frac{X(t)}{w(t)}\right)\!\right] - \frac{w(t)}{3}\!\left[ 11 + 15\textrm{csch}^{2}\!\left(\frac{X(t)}{w(t)}\right)\!\right]\!\right]\! ,
\label{dndt}
\end{eqnarray}
where we have used the fact that $\int_{-\infty}^{\infty} u_{xx} dx=0$. Let us now consider $\sigma^{2}(t)$, for which we get 
\begin{eqnarray}
\frac{d\sigma^{2}}{dt} &=& \frac{d}{dt}\left(\frac{1}{n(t)}\int_{-\infty}^{\infty} x^{2}u\, dx \right) = -\frac{1}{n^{2}(t)}\frac{dn}{dt}\int_{-\infty}^{\infty} x^{2}u\, dx + \frac{1}{n(t)}\int_{-\infty}^{\infty} x^{2}u_{t} dx \nonumber\\
&=& -\frac{\sigma^{2}(t)}{n(t)}\frac{dn}{dt} + \frac{1}{n(t)}\int_{-\infty}^{\infty} x^{2}\left[ Du_{xx}+ \rho u(1-u) \right]dx \nonumber\\
&=& -\frac{\sigma^{2}(t)}{n(t)}\frac{dn}{dt} + 2D + \rho\sigma^{2}(t) - \frac{\rho}{n(t)}\int_{-\infty}^{\infty} x^{2}u^{2}\, dx \nonumber\\
&=& 2D + \rho A(t)\frac{\left[ X(t)\coth\!\left(\frac{X(t)}{w(t)}\right)\!\!\left[ \frac{5X^{2}(t)}{3w(t)} + 7w(t) - 6X(t)\coth\!\left(\frac{X(t)}{w(t)}\right)\!\right] - w^{2}(t)\right]}{3\!\left[ \frac{X(t)}{w(t)}\coth\!\left(\frac{X(t)}{w(t)}\right) - 1 \right]^{2}},
\label{dsigma2dt}
\end{eqnarray}
where we have made use of Eqs. \eqref{numberint}, \eqref{varianceint} and \eqref{dndt}.
\par
Finally, for $\gamma(t)$, we obtain
\begin{multline}
\frac{d\gamma}{dt}  = \frac{d}{dt}\left(-\int_{0}^{\infty} u_{x} dx \right) = -\int_{0}^{\infty} u_{xt} dx = -\int_{0}^{\infty} \left[ Du_{xxx}+ \rho u_{x} - 2\rho u\, u_{x}) \right]dx \\
 = A(t)\tanh^{2}\left(\frac{X(t)}{2w(t)}\right)\!\left[ \rho - \rho A(t)\tanh^{2}\left(\frac{X(t)}{2w(t)}\right) - \frac{D}{w^{2}(t)}\textrm{sech}^{2}\left(\frac{X(t)}{2w(t)}\right)\right]\! .
\label{dgammadt}
\end{multline}
\par
Now, in order to get a  system of ordinary differential equations for the dynamically relevant quantities $A(t)$, $X(t)$ and $w(t)$ we can 
 differentiate Eqs. \eqref{numberint}-\eqref{gammaint} and use Eqs. \eqref{dndt}-\eqref{dgammadt}. Long calculations lead to a final closed set of ordinary differential equations 
that are written in Appendix \ref{ApA}. 
 
\par

\begin{figure}
\begin{center}
\includegraphics[scale=0.42]{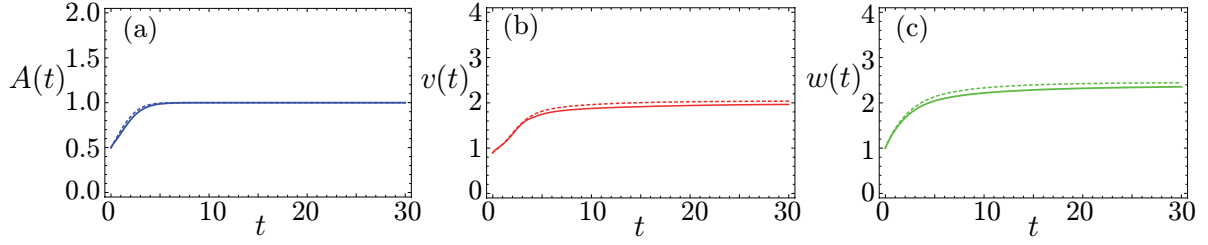}
\caption{[Color Online]. Comparison of the evolution of front solutions of the Fisher-Kolmogorov equation described by Eq. (\ref{dimensionalFK}) with 
the solutions of the effective particle method given by  Eqs. \eqref{EDOn}-\eqref{EDOgamma}. The FK equation is solved numerically using a standard second order in time finite difference method with zero derivative boundary conditions and constants $D=1$ and $\rho=1$. The initial data is given by Eq. (\ref{profilelocalized}) with  $A_0 = 0.5$, $w_0 =1$, and $X_0 =2$. The corresponding ODEs are solved taking the later as initial values for $A(t), w(t), X(t)$. In all subplots (a)-(c) the solid lines correspond to the parameters $A_{\text{PDE}}(t)$, $v_\text{PDE}(t)$ and $w_\text{PDE}(t)$,  extracted from the numerical solution of the FK equation and Eqs. \eqref{numberint}-\eqref{gammaint}. The dashed lines correspond to the results of the ODEs. The subplots shown are: (a) amplitude $A(t)$ (dashed) versus $A_{\text{PDE}}(t)$ (solid), (b) velocity of the front $v(t)$ (dashed) versus $v_{\text{PDE}}(t)$ (solid), (c) width of the solution $w(t)$ (dashed) versus $w_{\textrm{PDE}}(t)$ (solid).
\label{figura3}}
\end{center}
\end{figure}

\begin{figure}
\begin{center}
\includegraphics[scale=0.48]{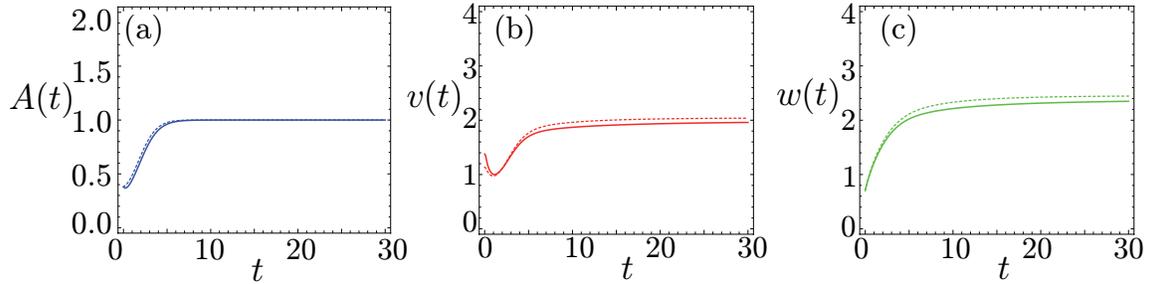}
\caption{[Color Online]. Comparison of the evolution of front solutions of the Fisher-Kolmogorov equation described by Eq. (\ref{dimensionalFK}) with 
the solutions of the effective particle method given by Eqs. \eqref{EDOn}-\eqref{EDOgamma}. The FK equation is solved numerically using a standard second order in time finite difference method with zero derivative boundary conditions and constants $D=1$ and $\rho=1$. The initial data is given by a Gaussian profile $u(x,0)=0.2e^{-0.5x^{2}}$. In all subplots (a)-(c) the solid lines correspond to the parameters $A_{\text{PDE}}(t)$, $v_\text{PDE}(t)$ and $w_\text{PDE}(t)$,  extracted from the numerical solution of the FK equation and Eqs. \eqref{numberint}-\eqref{gammaint}. The dashed lines correspond to the results of the ODEs. The figures shown are: (a) amplitude $A(t)$ (dashed) versus $A_{\text{PDE}}(t)$ (solid), (b) velocity of the front $v(t)$ (dashed) versus $v_{\text{PDE}}(t)$ (solid), (c) width of the solution $w(t)$ (dashed) versus $w_{\textrm{PDE}}(t)$ (solid).
\label{figura4}}
\end{center}
\end{figure}

\subsection{Comparison of the effective particle dynamics with the FK equation}

Let $X_0$, $A_0$ and $w_0$ as in Sec. denote the initial values of the right-front position, amplitude and width, respectively. To test the validity of Eqs. (\ref{EDOn}-\ref{EDOgamma}) 
as approximations to the full PDE dynamics for localized initial data, we have run extensive series of simulations for different parameter values. As in the case of the front solutions discussed in Sec. \ref{method} we have found that the effective particle model approximates with a very good accuracy the dynamics of Eqs. (\ref{dimensionalFK}).

As an example in Fig.  \ref{figura2} we compare the profiles of the numerical solution to the FK equation (\ref{dimensionalFK}) with the dynamics provided by the ansatz (\ref{profilelocalized}) together with Eqs. (\ref{EDOn}-\ref{EDOgamma}) for various times and different initial conditions. Figure \ref{figura2}(a) corresponds to the case where $X_0>w_0$, whereas in Fig. \ref{figura2}(b) $X_0<w_0$. Notice that in Fig. \ref{figura2}(b) the right and left fronts advance at a faster pace when compared to Fig. \ref{figura2}(a) despite the fact that the initial profile is much smaller. In both cases we get an excellent agreement since the profile chosen closely resembles the one arising spontaneously from the partial differential equation (\ref{dimensionalFK}). 

To get a more direct comparison between the parameter evolution computed from the ODEs (\ref{EDOn}-\ref{EDOgamma})  and the PDE (\ref{dimensionalFK}) we have compared also the evolution of the parameters in many simulations.  Figures \ref{figura3} and \ref{figura4} provide two typical examples. In Fig. \ref{figura3} the initial condition is given by the ansatz (\ref{profilelocalized}), whereas in Fig. \ref{figura4} the initial condition is a Gaussian profile that does not have initially the expected exponential decay of $u(x,t)$ for large values of $x$. Despite this deviation of the initial data  there is generally a very good agreement of the approximate effective particle equations with the simulations of the PDEs. \par

\subsection{Asymptotic regime}

Despite Eqs. (\ref{EDOn}-\ref{EDOgamma}) are ordinary differential equations allowing to get the evolution of the parameters in a more direct way than the PDE, the fact that they have a complicated structure originated in the interactions between the two fronts used to approximate the solution, makes its qualitative analysis complicated. However, there are several limits in which it is possible to elucidate the dynamics of $u(x,t)$ in terms of simpler expressions.
\par
Let us consider the case when the population density $u(x,t)$ is much more extended than the infiltrative zone, i.e.  $X(t)\gg w(t)$, a situation that is always verified for long enough times. In that case, the corresponding system of ordinary differential equations reduces to 
\begin{subequations}
\label{ODEslocalized}
\begin{eqnarray}
\frac{dA}{dt} & = & \rho A(t)\left[1-A(t)\right]\! , \label{Amplocalized} \\
\frac{dX}{dt} - \frac{dw}{dt}& = & \frac{5}{6}\rho A(t)w(t), \label{widthlocalized} \\
\frac{dX}{dt} - \frac{\pi^{2}}{3}\frac{dw}{dt}& = & \rho A(t)w(t) - \frac{D}{w(t)}\, . \label{positionlocalized} 
\end{eqnarray}
\end{subequations}
Notice that Eqs. \eqref{Amplocalized} and  \eqref{widthlocalized} are exactly the same as the first two equations in \eqref{ODEs}. Combining Eqs. \eqref{widthlocalized} and \eqref{positionlocalized} we arrive at the third equation in \eqref{ODEs}. This is consistent with the intuitively expected fact that if $u(x,t)$ becomes very broad then the left and right fronts no longer interact and behave as independent single-propagating particles whose time evolution is described by Eqs. \eqref{ODEs}. It is clear also that in the limit $t\to\infty$, the parameters $A(t)$, $w(t)$ and $v(t)$ tend to the asymptotic values given by Eqs. \eqref{asymptoticparameters}. Therefore, if $X(t)\gg w(t)$ the dynamics of the localized profile is the same as the simpler single front wave. 



\section{Applications to brain tumor dynamics (I): Transition to malignancy and time of birth of low grade gliomas}
\label{sec-apl}

\subsection{Motivation}

Low grade glioma (LGG) is a term used to describe World Health Organization grade II primary brain tumors of astrocytic and/or oligodendroglial origin \cite{WHO}.
These tumors are highly infiltrative and generally incurable but have a median survival time of $> 5$ years because of low proliferation \cite{Pignatti2002,Ruiz2009,Pouratian2010}.
While most patients remain clinically asymptomatic besides seizures, the tumor transformation to aggressive high grade glioma is eventually seen in most patients. 

Management of LGG has historically been controversial because these patients are typically young, with few, if any, neurological disorders. Historically, a wait and see approach was often favored in most cases of LGG, due to the lack of symptoms in these mostly young and otherwise healthy adults. The support for this practice came from several retrospective studies showing no difference in outcome (survival, quality of life) if therapy was deferred \cite{Olson2000,Batchelor2006}. Other investigations have suggested a prolonged survival through surgery \cite{Smith2008}. In absence of a randomized controlled trial, recently published studies may provide the most convincing evidence in support of an early surgery strategy \cite{Jakola2012} and waiting for the use of other therapeutical options such as radiotherapy and chemotherapy. However, the decision on the individual treatment strategy is based on a number of factors including patient preference, age, performance status, and location of tumor \cite{Ruiz2009,Pouratian2010}.

The FK equation arises as a basic model of the dynamics of low grade brain tumors, accounting in a simple way for the two main features of tumor cells: infiltration and proliferation (this type of tumors do not metastasize to other organs). In this context the density $u(x,t)$ describes a wave of invasive tumor cells as it has been studied in many papers \cite{SS1,SS2,SS3,Badoual}. One of the main characteristics of those tumors is their low cellular density. However, when the cellular density becomes too high, hipoxia arises triggering the hypoxic response \cite{SWCR2,PGRT} with a cascade of metabolic alterations in the cell, including genetic instability, and has a potential effect on the acceleration of the progression which may have an impact in what is called the transition to malignancy.  

\subsection{Estimates for the time of transition to malignancy}

Assuming that one of the events determining the transition to malignancy is the high local tumor density, we can get an estimate of the time taken by the tumor cell density to progress from an initial maximal density $A_0$ to a critical threshold density $A_*$ beyond which hypoxia and tissue damage trigger the cell changes leading to the malignant transformation. 

The exact analytical solutions provided by the effective particle method allow us to provide an estimate for the time required for that process. Finding this time $t_{*}$ from the FK generally requires numerical computation. In contrast, it is straightforward to get it from Eq. \eqref{Adet}. 

\begin{equation}\label{TTm}
T_{*}=\frac{1}{\rho}\log\left[\frac{(1-A_0)A_{*}}{(1-A_{*})A_0}\right].
\end{equation}

It is interesting to note that according to Eq. (\ref{TTm}), the time of transition to malignancy does not depend on the diffusion coefficient $D$. Eq. (\ref{TTm}) should be taken as an estimate or order of magnitude upper bound since tumor density profiles are not expected to be as smooth as our ansatz functions (e.g. Eq. \eqref{profile}) and 
the transition to malignancy may depend on the highest initial cell density present throughout the tumor. 

To use \eqref{TTm} in clinical scenarios it is necessary to estimate the parameters $A_0$, $A_*$ and $\rho$. Firstly, $A_0$ would correspond to the detection amplitude and has been discussed in several papers (see e.g. \cite{Swanson1} and references therein), $A_*$ is less well known but can be estimated to be in the range between 0.5 and 0.8. However more work with real data is necessary to confirm this choice. Finally the estimate of $\rho$ for individual patients is very difficult in the case of low-grade gliomas because of the slow and irregular growth of these tumors. Recent work has suggested that this number can be obtained from the response of the tumor to radiotherapy \cite{PGRT} and it seems reasonable that those estimates may correlate also with the histology results for proliferation markers when available (e.g. MIB/Ki-67 labelling index).

\subsection{Estimates for the time of birth of the tumor}

Running the equations (\ref{EDOn}-\ref{EDOgamma}) backwards in time it is also possible to propose an estimate for the time of birth of the tumor, i.e. the time for which the tumor maximum density corresponds to a single cell $A_s$ (assuming that the tumor starts from one mutated cell). In our case, since astrocytic cells have a typical size of 10 $\mu$m and thus the maximal linear density is about 100 cells/mm, we would get $A_s \simeq 0.01$. The equation 
\begin{equation}\label{TTi}
T_{s}=\frac{1}{\rho}\log\left[\frac{(1-A_s)A_{0}}{(1-A_{0})A_s}\right],
\end{equation}
gives the tumor time of birth. As with Eq. (\ref{TTm}), Eq. (\ref{TTi}) depends inversely on $\rho$ and thus, assessing this parameter from the available patient's data may result in a value dependent on each patient. Getting estimates for $T_s$ is very relevant clinically in order to correlate the prediction with the patient's clinical history and to extract information on possible causes for the origin of this type of tumors. Up to now, they are thought to be sporadic, but there is a strong interest among clinicians to investigate possible causes for these tumors, for which Eq. (\ref{TTi}) may be helpful. In fact, this problem has been considered in the framework of simulations of the full FK equation in \cite{Badoual}. Our approach complements that work providing an equation that allows to circumvent the direct simulation of the PDE. 

\section{Applications to brain tumor dynamics (II): Simple description of the response to radiotherapy}
\label{sec-apl2}

\subsection{Radiotherapy of low grade gliomas}

The effectiveness of radiation therapy against low grade gliomas was proven in the clinical trial of Ref. \cite{trial6}. However, the timing of radiotherapy after biopsy or debulking is debated. It is now well known that immediate radiotherapy after surgery increases the time of response to the therapy known as progression-free survival, but does not seem to improve the overall survival time. At the same time, radiotherapy may contribute to the observed neurological deficit of this patients as a result of the damage to the normal brain \cite{VandenBent2005}. This is why radiotherapy is usually deferred in time until an increase in the tumor grade is diagnosed or, else, offered to selected patients with a combination of low risk factors such as age, sub-total resection, and diffuse astrocytoma pathology \cite{trial5}. 

It is interesting that slight modifications of the radiotherapy protocol have been found to have little or no effect on overall survival \cite{Karim}.  
Mathematical modelling has the potential to select patients that may benefit from early radiotherapy. Also, it may help in developing specific optimal fractionation schemes for selected patient subgroups. However, despite its enormous potential, mathematical modelling has had a very limited use with strong focus on some aspects of radiation therapy (RT) for high-grade gliomas \cite{Powatil2007,Rockne2010,BondiauRT,Konokoglu2010,Kirkby2010,Stamatakos2006}. Up to now, no ideas coming from mathematical modelling have been found useful for clinical application in any of these contexts.

There is thus a need for models accounting for the fundamental features of low-grade glioma dynamics and their response to radiation therapy without involving excessive details on the -often unknown- specific processes.  The increasing availability of systematic and quantitative measurements of LGG growth rates provides key information for the development and validation of such models \cite{Pallud1,Pallud2}.

\subsection{A simple mathematical model of tumor response to therapy}

 We will assume that the radiation doses $d_k$ (Gy) are given instantaneously at times $t_k$ for integer $k=1,...,n$, what implies assuming that the total irradiation time is very short in comparison with the tumor response times. This is typically the case in external beam radiotherapy where treatment times are in the order of minutes while tumor response times are in the range of weeks or months. We will also assume that the radiation is spatially uniform through the tumor what is also a good approximation to clinical practice whenever possible. Following the standard practice in radiotherapy, we will take the damaged fraction of tumor cells as given by the classical linear-quadratic (LQ) model \cite{Joiner2009}, i.e. the fraction of cells that are not lethally damaged by a dose $d_k$, to be given by
\begin{equation}\label{LQ}
\text{SF}_{d_k}=e^{\displaystyle{-\alpha_t d_k -\beta_t d_k^{2}}}, 
\end{equation} 
where $\alpha_t$ $(\text{Gy}^{-1})$ and $\beta_t$  $(\text{Gy}^{-2})$ are respectively the linear and quadratic coefficients for \emph{tumor} cell damage of the LQ model. The precise values of the parameters remain unknown despite many studies, because what is clinically more relevant is the ratio $\alpha_t/\beta_t$ which for glioma cells is around 10. The full treatment consists of a total dose $D$ split in a series of -typically equal- $n$ doses $d_j$ delivered at times $t_j$. For instance, for high grade gliomas the standard protocol consists of $n=30$ doses of $d = 2$ Gy for a total of $D=$ 60 Gy, administered once per day (except for the weekends) during 6 weeks. For LGGs typical doses range from 45 to 54 Gy with $d=1.8 $ Gy during 5 or 6 weeks of treatment. 



Taking into account all this information, and depending on the cell death mechanism, different models can be written. In Ref. \cite{PGRT} a model has been proposed including delayed response to radiation. However the simplest possible model can be based on the 
assumption that damaged cells die instantaneously after the therapy (or in times very short compared with the characteristic natural history of the tumor. This leads to the simplest possible model where the evolution of the tumor cell density is given by Eq. (\ref{dimensionalFK}) or its ODE effective particle reduction described by \eqref{ODEs}) for each interval between doses $[t_k,t_{k+1}]$. The initial data for each subinterval will be take then as given by 
\begin{eqnarray}
u(0,x) & = & u_0(x), \\
u(t_k^+,x) & = & \text{SF}_{d_k} u(t_k,x),
\end{eqnarray}
where $u(t_k^+,x)$ is the density after the dose $k$ is given.
 
 \begin{figure}
\centering
\epsfig{file=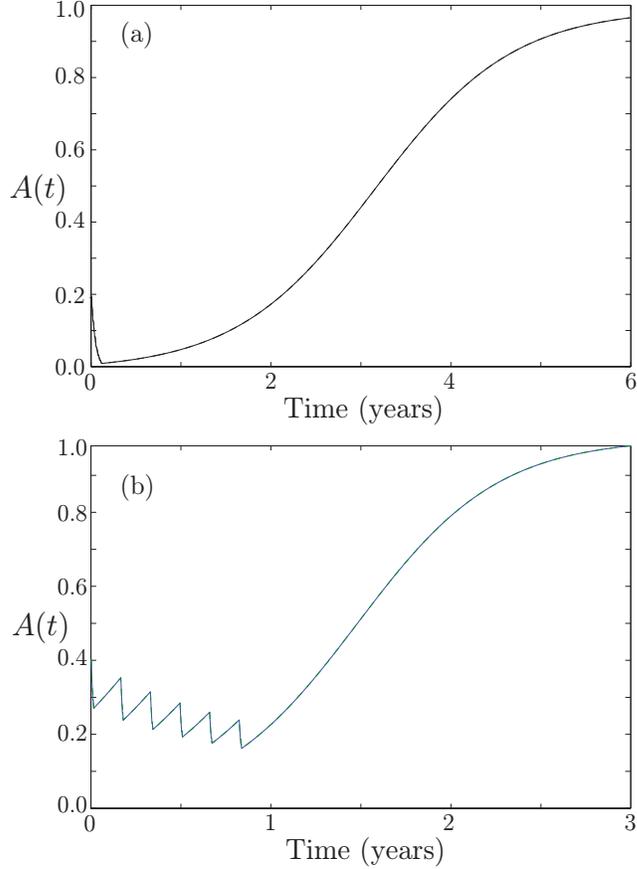,width=0.60\textwidth}
\caption{Tumor amplitude evolution under radiation therapy for several choices of the parameter values. In both cases the solid line correspond to the amplitude computed using the FK equation \eqref{dimensionalFK} using the equation $A_{FK}(t)= \max_x\left(C(x,t)\right)$ and the dashed line (overlapping with the solid one) corresponds to the approximation of the effective particle method given by Eqs. (\ref{recur}) and (\ref{At}). Shownare: (a) $\rho = 0.00356$ day$^{-1}$,  $ D= 0.0075$ mm$^2$/day, and initial tumor cell density given by $C(x,0) = 0.2 /\left[1-0.2+0.2 \exp\left(x/15\right)\right]$ with $x$ measured in mm. Radiotherapy follows the standard scheme (6 weeks with 1.8 Gy doses from monday to friday) and starts at time $t=7$ days. The differences between both curves are minimal with  $\|A(t) - A_{FK} \|_{\infty} = 0.0014$. (b) $\rho = 0.007$ day$^{-1}$,  $ D= 0.0075$ mm$^2$/day, and initial tumor cell density given by $C(x,0) = 04 /\left[1-0.4+0.4 \exp\left(x/15\right)\right]$ with $x$ measured in mm. Radiotherapy follows a non-standard scheme with radiation sessions the first week of every two months for a total of 6 weeks with 1.8 Gy doses from monday to friday and starts at time $t=1$ day.  The differences between both curves are minimal with  $\|A(t) - A_{FK} \|_{\infty} = 0.0025$.
 \label{RTT}}
\end{figure}

This allows us to write a simple model for the response of the tumor amplitude to radiation given by Eq. (\ref{LQ}) together with Eq. (\ref{Adet}). For instance, starting from an initial amplitude $A(t_0)$ at time $t_0$, we get that before and after the first irradiation at time $t_1$ the tumor amplitudes will be given by
\begin{subequations}
\begin{eqnarray*}
A(t^-_1) & = & \frac{A(t_0) e^{\rho (t_1-t_0)}}{1+ A(t_0) \left[e^{\rho (t_1-t_0)}-1\right]}, \\
A(t^+_1) & = &  \frac{\text{SF}_k A(t_0) e^{\rho (t_1-t_0)}}{1+ A(t_0) \left[e^{\rho (t_1-t_0)}-1\right]}, 
\end{eqnarray*}
\end{subequations}
Thus, defining $\hat{A}_{k} \equiv A(t_k^+)$ and taking by definition $\hat{A}_0 = A(t_0)$ we arrive to the following recursive
 formula for the tumor amplitudes after each irradiation cycle 
\begin{equation}\label{recur}
\hat{A}_{k+1}  =   \frac{\text{SF}_{k} \hat{A}_k e^{\rho (t_{k+1}-t_k)}}{1 + \hat{A}_k \left[ e^{\rho (t_{k+1}-t_k)}-1\right]}.
\end{equation}
Thus, the amplitude $A(t)$ of the tumor after a sequence of radiation sessions administered at times $(t_1, ..., t_n)$ with survival fractions $(\text{SF}_1, ..., \text{SF}_n)$, in the framework of our simple model, is given for all times by the equations 
\begin{subequations}
\label{At}
\begin{eqnarray}
A(t) &  =   & \frac{\hat{A}_k e^{\rho (t-t_k)}}{1 + \hat{A}_k \left[ e^{\rho (t-t_k)}-1\right]}, \ \ t \in (t_k,t_{k+1}],  \\
A(t) &  =   & \frac{\hat{A}_n e^{\rho (t-t_n)}}{1 + \hat{A}_n \left[ e^{\rho (t-t_n)}-1\right]}, \ \ t \geq t_n,
\end{eqnarray}
\end{subequations}
where the constants $ \hat{A}_k$, $k=1,..., n$ are obtained recursively using Eq. (\ref{recur}). 

\subsection{Validation and implications}

Eqs. (\ref{recur}) and (\ref{At}) provide a very simple model of the response of a low grade glioma to radiation therapy. However these simple formulae have been obtained in the framework of the simple effective particle method. To test if the approximation provided by this approach is acceptable when describing the solutions of Eq. (\ref{dimensionalFK}) together with an instantaneous response to radiation given by Eq. \eqref{LQ} we have numerically simulated these equations, computed the amplitude using $A(t) = \| u(x,t) \|_{\infty}$. The evolution of this ``exact" amplitude has been compared with the approximation provided by Eqs. (\ref{recur}) and (\ref{At}) obtained in the framework of the effective particle approximation to the dynamics.

We have compared the evolution in different scenarios. The first one corresponds to the standard radiation fractionation of a total of 54 Gy in 30 fractions of 1.8 Gy over a time range of 6 weeks (5 sessions per week from monday to friday).  Radiation is started 1 week after the initial simulation time ($t=0$) and the evolution is followed for six years. The comparison between the approximation based on the effective particle method and the full Fisher-Kolmogorov equation is displayed in Fig. \ref{RTT}(a). A second example is provided in Fig. \ref{RTT}(b) where a faster growing tumor ($\rho = 0.007$ day$^{-1}$ versus $\rho = 0.00356$ day$^{-1}$ in the previous case) is irradiated following a non-standard radiotherapy scheme consisting of five consecutive sessions of 1.8 Gy every two months with the intention to control the tumor rather than focusing on minimizing the tumor mass. The agreement between both approaches is excellent. 
\par

We have explored several parameter regimes finding a very good accuracy in the approximation provided by the effective particle method meaning that one can follow the evolution of the tumor amplitude using the simple equations Eqs. (\ref{recur}) and (\ref{At}). This approach allows us to reduce the problem of solving a partial differential equation to computing a finite number of iterations of a simple nonlinear mapping. This fact has very interesting implications since one may then pose optimization problems for finding optimal fractionation schemes in a much simpler context. One example is optimizing radiation for delaying the transition to malignancy, a problem that can be reformulated completely in terms of amplitudes assuming that there is a critical amplitude $A_*$ beyond which the transition to malignancy is triggered. Thus the problem becomes a discrete optimization one with a very simple non-differential model that is equivalent to finding the extrema of a function of several variables $(t_j, D_j)$ with several restrictions. While the complete analysis of this problem is beyond the scope of this paper and will be considered elsewhere, we want to highlight the potential of the effective particle method to obtain  simple equations amenable to a complete analysis.

\section{Extension to related Fisher-Kolmogorov equations}
\label{sec-V}

\begin{figure}
\begin{center}
\includegraphics[scale=0.27]{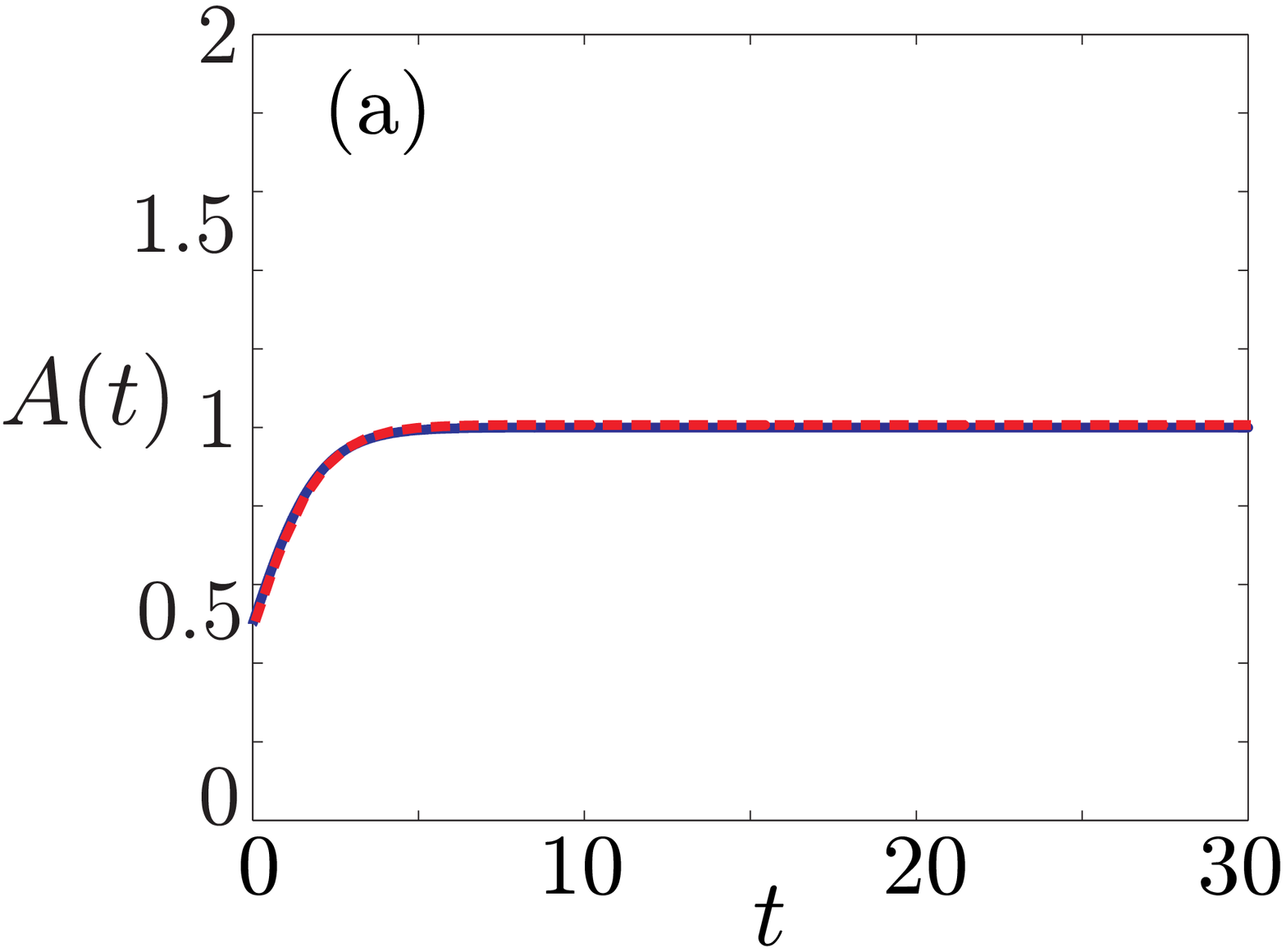}
\includegraphics[scale=0.27]{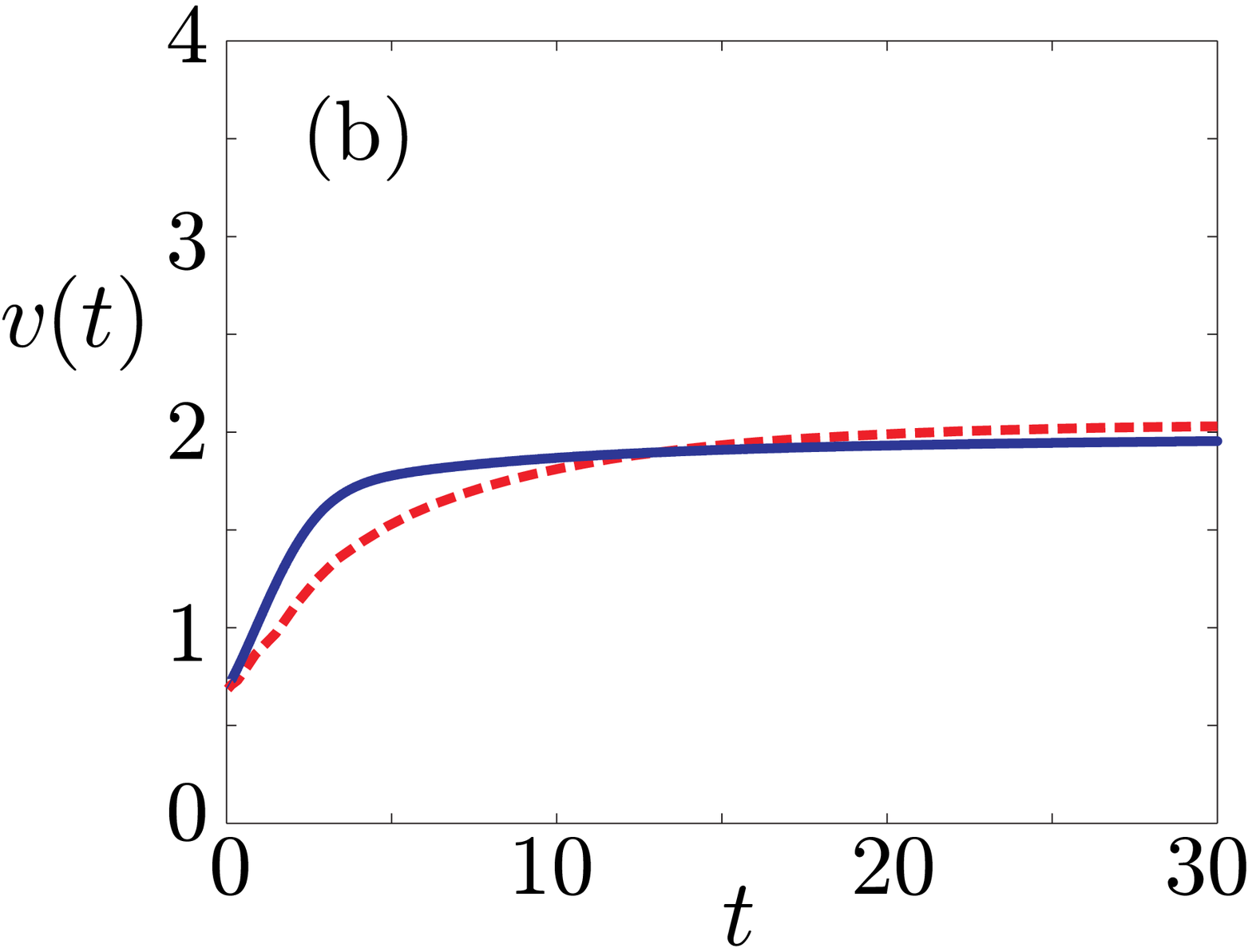}
\includegraphics[scale=0.27]{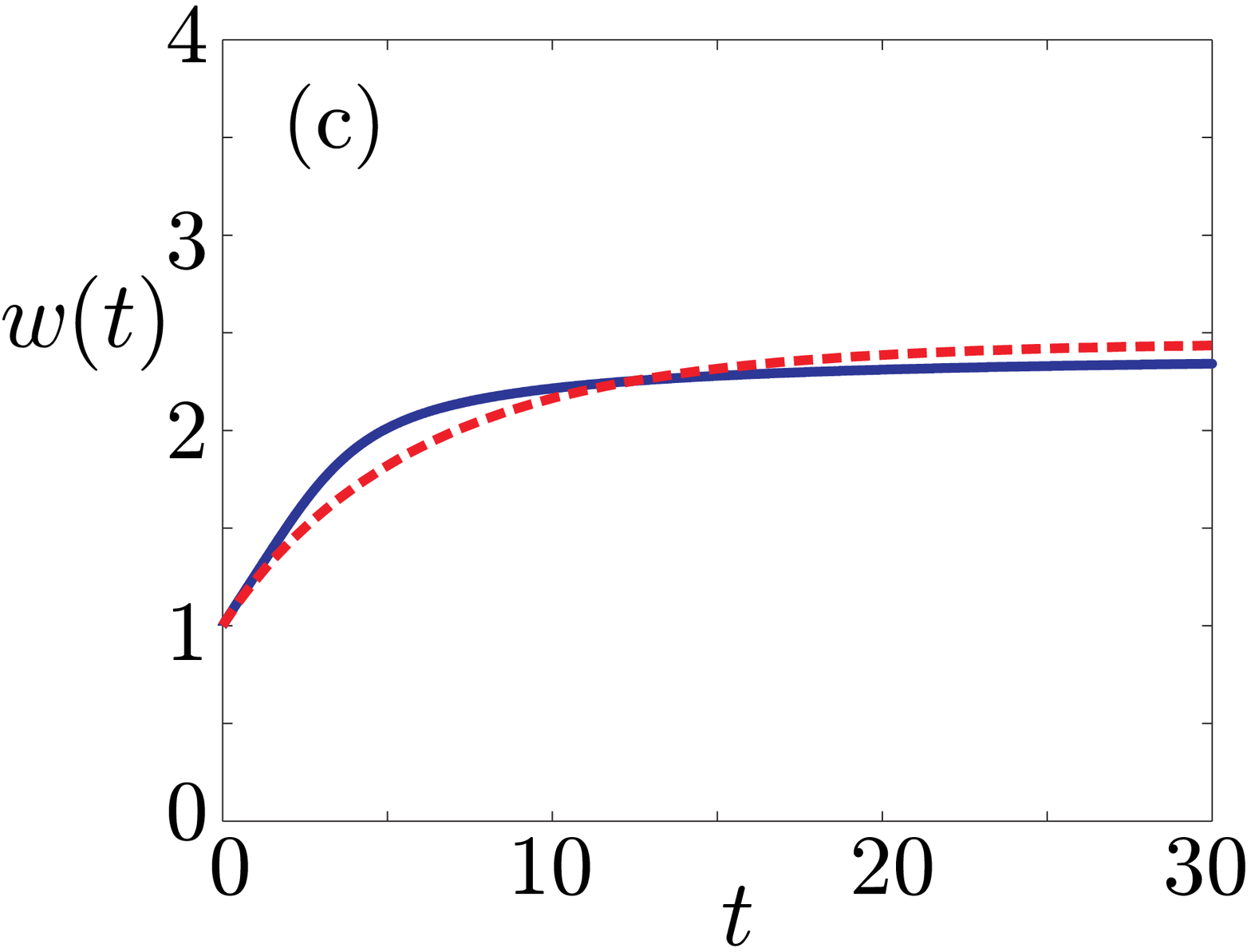}
\caption{[Color Online]. Comparison of the evolution of front solutions of the Fisher-Kolmogorov equation  with space-dependent diffusion described by Eq. (\ref{FKDx}) with 
the analytical solutions of the effective particle method given by Eqs. \eqref{ODEs}. The FK equation is solved numerically using a standard second order in time finite difference method with zero derivative boundary conditions with $D(x)$ as \eqref{Dwg}, where $D_{\text{w}}=1$, $D_{\text{g}}=0.5$, $\alpha=0$ and $\rho=1$. The initial data is given by Eq. (\ref{profile}) with  $A_0 = 0.5$, $w_0 =1$, and $X_0 =0.5$. In all subplots (a)-(c) the solid lines correspond to the parameters \eqref{paramsPDE} extracted from the numerical solution of the FK equation. The dashed lines correspond to the analytical solutions of the ODEs. The subplots show the: (a) amplitude $A(t)$ (dashed) versus $A_{\textrm{PDE}}(t)$ (solid), (b) velocity of the front $v(t)$ (dashed) versus $v_{\text{PDE}}(t)$ (solid), (c) width of the solution $w(t)$ (dashed) versus $w_{\textrm{PDE}}(t)$ (solid).
\label{figura6}}
\end{center}
\end{figure}

The effective particle method can be extended to other types of reaction-diffusion equations having fronts as asymptotic attractors of the dynamics. For instance, for the cases where the diffusion coefficient either density dependent $D=D(u)$, space and/or time dependent $D=D(x,t)$ and/or the growth rates also have extra dependencies $\rho=\rho(x,t)$, as well as for other situations where the system is multicomponent (e.g. in cancer modelling, there are several phenotypes). These examples are of interest in a broad range of biological processes \cite{Shigesada} and particularly in cancer modelling problems \cite{Maini,Ayache2,graywhite,Ayache1,JB}. Here, as an example of how this methodology is robust and can be extended beyond the most basic FK equation, we consider the equation with space-dependent diffusion
\begin{equation}\label{FKDx}
u_{t}=[D(x)u_{x}]_{x}+\rho u(1-u).
\end{equation}
Thus, we are considering Fickian diffusion  but with a diffusion coefficient dependent on the space coordinates \cite{Murray}. 
Then, we can find the set of effective particle equations for the amplitude $A$, the center of mass $X$ and the width $w$ of the travelling wave \eqref{profile}, for Eq. \eqref{FKDx} following the same procedure of Section \ref{method}. The result is
\begin{subequations}
\label{ODEsDx}
\begin{eqnarray}
\frac{dA}{dt} & = & \rho A(t)\left[1-A(t)\right]\! , \label{coco} \\
\frac{dX}{dt} - \frac{dw}{dt}& = & \frac{5}{6}\rho A(t)w(t), \label{cece} \\ 
\frac{dw^2}{dt}&=&-\frac{6}{(\pi^{2}-3)A(t)}\int_{-\infty}^{\infty}D(x)u_{x}-\frac{\rho}{\pi^{2}-3}A(t)w^{2}(t). \label{widthdx}
\end{eqnarray}
\end{subequations}
To test the limits of the method we will use a piecewise constant discontinuous diffusion coefficient given by
 \begin{equation}\label{Dwg}
 D(x)=\begin{cases}
 D_{\text{g}} & \mbox{if $x\leq \alpha$},\\
 D_{\text{w}} & \mbox{if $x>\alpha$}.
 \end{cases}
 \end{equation} 
 Effective particle methods have tipically difficultities in dealing with discontinuous or very fast varying coefficients, and other effects that influence strongly the shape of the solution thus providing asymmetric deformations of the initial ansatz. However as we will see in what follows, in this case the method is able to follow qualitatively the details of the evolution of the deformed front solution. 
 
 The explicit choice for the diffusion coefficient given by Eq. (\ref{Dwg}) has also some interest in applications. Specifically, in brain tumor modelling it arises in situations in which the tumor invades the gray matter from the white matter. In that case Eq. (\ref{FKDx}) has been proposed as a toy model in which the diffusion coefficient is a piecewise constant function, corresponding to different tumor cell motilities in the white and grey matter \cite{SS3,Ayache2,Murray}. Analogous mathematical problems arise in other application scenarios (see e.g. \cite{Gen2} and references therein). 
The explicit form of $D(x)$ given by Eq. (\ref{Dwg}) allows us to 
 compute explicitly the integral in 
Eq. \eqref{widthdx} to get
\begin{equation}\label{cucu}
\frac{dw^2}{dt}=\frac{6}{(\pi^{2}-3)} \left[ D_{\text{g}}-\frac{D_{\text{g}}-D_{\text{w}}}{\left[1+e^{(\alpha-X(t))/w(t)}\right]^{2}}\right]-\frac{\rho}{\pi^2-3}A(t)w^2(t).
\end{equation}
Eq. (\ref{cucu}) together with Eqs (\ref{coco}) and (\ref{cece}) is again a closed system of ODEs ruling the dynamics of the front in the framework of the effective particle method. We have run extensive simulations to compare the dynamics of Eqs. \eqref{ODEsDx} with the numerical solution calculated for Eq. \eqref{FKDx} and the ansatz \eqref{profile}. Despite the potential problems that might be expected coming from the discontinuity of the diffusion coefficient the agreement is very in all of the cases studied. A typical example is shown in Fig. \ref{figura6} where it is seen how the amplitude dynamics is fully captured by the effective particle method and the width and speed dynamics have only quantitative transient differences with the asymptotic behavior been again correctly described by the approximation method. 

\section{Conclusions}
\label{discussion}

In this paper we have presented an extension of effective particle methods to deal with a non-Hamiltonian problem of relevance in applied science: the Fisher-Kolmogorov (FK) equation. 
 The method provides a very simple picture in terms of ordinary differential equations of the behavior of both a single-front and a localized travelling wave. 
 It yields direct information on three relevant parameters: the amplitude, the front position and the width of the wave, which turn out to be parameters more easily accessible to experimental measurement in application scenarios
  than the entire profile $u(x,t)$, yet furnishing sufficient insight on the characteristic dynamics of  the dynamics of the partial differential equation in certain regimes. 

\par 
 
In addition to presenting the method and quantifying its accuracy, we have also discussed, through the specific application to problems arising in the description of brain tumors, how it can be used to get very simple estimates useful for applied scientists. Specifically we have developed explicit formulae to estimate the times of transition to malignancy and of birth of a low grade glioma. We have also provided a  way to transform the problem of optimizing radiation delivery on the PDE to a finite-dimensional problem involving only a discrete map.

\par

The method presented in this paper has very broad implications and potential uses. As an example we have shown its appropriateness to deal with a problem with spatially dependent diffusion with discontinuous diffusion coefficient. However,  it can be extended to many other reaction diffusion equations in order to  get a simple qualitative understanding of the dynamics of coherent structures. Secondly extending it to higher dimensions, may allow to get simple models in situations were theoretical results are much more scarce and numerical simulations more difficult. In that case approximate front profiles may be used as tentative test functions for the method \cite{SIAMTW1}. Finally, the set of ODEs provided by the effective particle method also allow simplifying optimal control problems, such as those involved in finding the optimal combinations of different therapies, that are much more difficult to cope within the framework of partial differential equations.

\par

We hope that this paper would further stimulate the application of the method to get useful information for the many applications of the Fisher-Kolmogorov and related equations. 

\section*{Acknowledgements} 

This work has been supported by grants MTM2009-13832 and MTM2012-31073 (Ministerio
de Econom\'{\i}a y Competitividad, Spain).
We would like to acknowledge Alicia Mart\'{\i}nez (Universidad de Castilla-La Mancha, Spain) and Philip Maini (Oxford University, UK) for discussions.

\appendix 

\section{Full form of the effective particle equations for localized initial data}
\label{ApA}

For completeness, we detail the full expressions of the system of ordinary differential equations for $A(t)$, $X(t)$ and $w(t)$ which follow by differentiating Eqs. \eqref{numberint}-\eqref{gammaint} and equating them to Eqs. \eqref{dndt}-\eqref{dgammadt}. 
\par
The first equation corresponds to the total number $n(t)$
\begin{eqnarray}
&2&\!\!A(t)\!\left[ \coth\!\left(\frac{X(t)}{w(t)}\right)\!\frac{dX}{dt} - \frac{dw}{dt} - \textrm{csch}^{2}\!\left(\frac{X(t)}{w(t)}\right)\!\left(\frac{X(t)}{w(t)}\frac{dX}{dt}- \frac{X^{2}(t)}{w^{2}(t)}\frac{dw}{dt}\right)\right] \nonumber\\
&&+\, 2\left(\frac{dA}{dt} - \rho A(t)\right)\!\left[ X(t)\coth\!\left(\frac{X(t)}{w(t)}\right) - w(t)\right] \nonumber\\
&=& \rho A^{2}(t)\!\left[ X(t)\coth\!\left(\frac{X(t)}{w(t)}\right)\!\!\left[ 2 + 5\textrm{csch}^{2}\!\left(\frac{X(t)}{w(t)}\right)\!\right] - \frac{w(t)}{3}\!\left[ 11 + 15\textrm{csch}^{2}\!\left(\frac{X(t)}{w(t)}\right)\!\right]\!\right]\! .
\label{EDOn}
\end{eqnarray}
\par
The second equation corresponds to the variance $\sigma^{2}(t)$
\begin{eqnarray}
&& \hspace*{-5mm}\frac{2X^{2}(t)\!\left[ w(t)\coth\!\left(\frac{X(t)}{w(t)}\right)\!\frac{dX}{dt} - X(t)\coth\!\left(\frac{X(t)}{w(t)}\right)\!\frac{dw}{dt} -X(t)\textrm{csch}^{2}\!\left(\frac{X(t)}{w(t)}\right)\!\left( \frac{dX}{dt} - \frac{X(t)}{w(t)}\frac{dw}{dt}\right)\right]}{3\!\left[ X(t)\coth\!\left(\frac{X(t)}{w(t)}\right) - w(t)\right]^{2}}\nonumber\\
\hspace*{5mm} &+& \frac{2}{3}X(t)\frac{dX}{dt} + \frac{2\pi^{2}}{3}w(t)\frac{dw}{dt} - \frac{4X(t)w(t)}{3\!\left[ X(t)\coth\!\left(\frac{X(t)}{w(t)}\right) - w(t)\right]}\frac{dX}{dt}\nonumber \\
\hspace*{5mm} &=& 2D + \rho A(t)\frac{\left[ X(t)\coth\!\left(\frac{X(t)}{w(t)}\right)\!\!\left[ \frac{5X^{2}(t)}{3w(t)} + 7w(t) - 6X(t)\coth\!\left(\frac{X(t)}{w(t)}\right)\!\right] - w^{2}(t)\right]}{3\!\left[ \frac{X(t)}{w(t)}\coth\!\left(\frac{X(t)}{w(t)}\right) - 1 \right]^{2}} \, .
\label{EDOsigma2}
 \end{eqnarray}
\par
Finally, the third equation corresponding to the right-front size $\gamma(t)$ is 
\begin{multline}
\tanh^{2}\!\left(\frac{X(t)}{2w(t)}\right)\!\frac{dA}{dt} + \frac{A(t)}{w(t)}\,\textrm{sech}^{2}\!\left(\frac{X(t)}{2w(t)}\right)\!\tanh\!\left(\frac{X(t)}{2w(t)}\right)\!\left(\frac{dX}{dt}- \frac{X(t)}{w(t)}\frac{dw}{dt}\right) \\
= A(t)\tanh^{2}\!\left(\frac{X(t)}{2w(t)}\right)\!\left[ \rho - \rho A(t)\tanh^{2}\!\left(\frac{X(t)}{2w(t)}\right) - \frac{D}{w^{2}(t)}\textrm{sech}^{2}\!\left(\frac{X(t)}{2w(t)}\right)\right]\! .
\label{EDOgamma}
\end{multline}

\newpage


\begin{thebibliography}{1}

\bibitem{General1} T. Dauxois, M. Peyrard, \emph{Physics of Solitons}, Cambridge University Press, 2006.

\bibitem{General2} A. Scott, \emph{Nonlinear Science: Emergence and Dynamics of Coherent Structures}, Oxford University Press, 2003.

\bibitem{General3} P. G. Drazin, R. S. Johnson, \emph{Solitons: An introduction}, Cambridge University Press, 1989.

\bibitem{General4} J. Yang, \emph{Nonlinear Waves in Integrable and Non-integrable Systems}, SIAM, 2010.

\bibitem{Kivshar1} Y. S. Kivshar, B. A. Malomed, \emph{Dynamics of solitons in nearly integrable systems}, Rev. Mod. Phys. 61, 763-915 (1989).

\bibitem{Var1} B. A. Malomed, \emph{Variational methods in nonlinear fiber optics and related fields}, \emph{Prog. Opt.} 43, 7-193 (2002). 

\bibitem{Var2} N. R. Quintero, E. Zamorano-Sillero, \emph{Lagrangian formalism in perturbed nonlinear Klein-Gordon equations}, Physica D 197, 63 (2004).

\bibitem{Var3}V. M. P\'erez-Garc\'{\i}a, \emph{Self-similar solutions and collective coordinate methods for nonlinear Schr\"odinger equations}, Physica D 191, 211-218 (2004).

\bibitem{Var4} R. Carretero-Gonz\'alez, D. J. Frantzeskakis and P. G. Kevrekidis, \emph{Nonlinear waves in BoseÐEinstein condensates: physical relevance and mathematical techniques}, Nonlinearity 21, R139 (2008).

\bibitem{Var5} V. M. P\'erez-Garc\'{\i}a, P. Torres, G. D. Montesinos, \emph{The method of moments for Nonlinear ScheÁr\"odinger equations: Theory and Applications}, SIAM J. Appl. Math. 67, 990 (2007).

\bibitem{Var6} S. Cuenda. A. S\'anchez, \emph{Length scale competition in nonlinear Klein-Gordon models: A collective coordinate approach}, Chaos 15, 023502 (2005).

\bibitem{Var7} N. R. Quintero, F. G. Mertens, A. R. Bishop, \emph{Generalized traveling-wave method, variational approach, and modified conserved quantities
for the perturbed nonlinear Schršdinger equation}, Phys. Rev. E 82, 016606 (2010).

\bibitem{Red4} C. Foias, B. Nicolaenko, R. Temman (Eds.), \emph{The connection between infinite dimensional and finite dimensional dynamical systems}, American Mathematical Society (1987).

\bibitem{Red1} T. Bohr, M. H. Jensen, G. Paladin, A. Vulpiani, \emph{Dynamical systems approach to turbulence}, Cambridge University Press, 1998.

\bibitem{Red2} R. Joly, G. Raugel, \emph{A striking correspondence between the dynamics generated by the vector fields and by the scalar parabolic equations},  Confluentes Mathematici \textbf{3},  471-493 (2011).

\bibitem{Red3} A. N. Carvalho, J. W. Cholewa, G. Lozada-Cruz, M. R. T. Primo, \emph{Reduction of infinite dimensional systems to finite dimensions: Compact convergence approach}, Cuadernos de Matem\'atica \textbf{11}, 281-330 (2010).

\bibitem{Murray} J. Murray, {\em Mathematical biology}, Third Edition, Springer (2007).
 
\bibitem{Shigesada} N. Shigesada, K. Kawasaki, \emph{Biological Invasions: Theory and Practice}, Oxford University Press (1997).

\bibitem{PP}V. Volpert, S. Petrovskii, \emph{Reaction-difusion waves in biology}, Physics of Life Reviews \textbf{6}, 267Ð310  (2009).

 \bibitem{Swanson1} K. R. Swanson, R. Rostomily, E. C. Alvord, Jr, \emph{Predicting Survival of Patients with Glioblastoma by Combining a Mathematical Model and Pre-operative MR imaging Characteristics: A Proof of Principle}, Br. J. Cancer, \textbf{98}, 113-119 (2008).
 
 \bibitem{Swanson2} C. Wang, J. K. Rockhill, M. Mrugala, D.L. Peacock, A. Lai, K. Jusenius, J. M. Wardlaw, T. Cloughesy, A. M. Spence, R. Rockne, E. C. Alvord Jr., K. R. Swanson, \emph{Prognostic significance of growth kinetics in newly diagnosed glioblastomas revealed by combining serial imaging with a novel biomathematical model}, Cancer Res. \textbf{69}, 9133-40 (2009).
 
 \bibitem{PG1} V. M. P\'erez-Garc\'{\i}a, G. F. Calvo, J. Belmonte-Beitia, D. Diego, L. A. P\'erez-Romasanta, 
 \emph{Bright solitons in malignant gliomas}, Phys. Rev. E \textbf{84}, 021921 (2011).

\bibitem{Genzer2007}  J.F. Douglas, K. Efimenko, D. A. Fischer, F. R. Phelan, J. Genzer, \emph{Propagating waves of self-assembly in organosilane monolayers}, Proc. Nat. Acad. Sci. \textbf{104}, 10324 (2007).

\bibitem{Gen3} I. Epstein and J. A. Pojman, \emph{An Introduction to Nonlinear Chemical Dynamics}, Oxford University Press, New York (1998).

\bibitem{Gen4} P. Grindrod, \emph{The Theory and Applications of Reaction-Diffusion Equations}, Oxford University Press, New York (1996).

\bibitem{Gen1} J. Xin, \emph{Front Propagation in Heterogeneous Media}, SIAM Rev., \textbf{42}, 161-230 (2000).

\bibitem{Gen2} C. W. Curtis, D. M. Bortz, \emph{Propagation of fronts in the Fisher-Kolmogorov equation with spatially varying diffusion}, Phys. Rev. E \textbf{86}, 066108 (2012).




\bibitem{KKK} A. Kolmogoroff, I. Petrovsky, and N. Piscounoff, \emph{Etude de l'equation de la diffusion avec croissance
de la quantite de matiere et son application a un probleme biologique}. Moscow University, Bull.
Math. \textbf{1}, 1-25 (1937).

\bibitem{Sherrat} J. A. Sherrat, \emph{On the transition from initial data to travelling waves in the Fisher-KPP equation}, Dynamics and Stability of Systems \textbf{13}, 167-174 (1998).

 \bibitem{Ablowitz} M. J. Ablowitz, A. Zeppetella, \emph{Explicit solutions of Fisher's equation for a special wave speed}, Bull. Math. Biol. \textbf{41}, 835  (1979).
 
\bibitem{WHO}
Louis, D. N., Ohgaki, H., Wiestler, O. D.,  Cavenee, W. K., Burger, P. C., Jouvet, A., Scheithauer, B. W. \& Kleihues P, \textit{World health organization classification of tumours of the central nervous system}, 4th ed., Renouf Publishing Co. Ltd., Geneva. pp. 33-46 (2007)

\bibitem{Pignatti2002} Pignatti, F., Van den Bent, M., Curran, D., Debruyne, C., Sylvester, R., Therasse, P., Afra, D., Cornu, P., Bolla, M., Vecht, C. \& Karim, A.B., \emph{Prognostic factors for survival in adult patients with cerebral low- grade glioma}, J. Clin. Oncol., \textbf{20}, 2076-84 (2002).

\bibitem{Ruiz2009} Ruiz, J., \& Lesser, G. J., \emph{Low-Grade Gliomas}, \emph{Curr. Treat. Opt. Oncol.}, \textbf{10}, 231-242 (2009).

\bibitem{Pouratian2010} Pouratian, N., \& Schiff, D.  \emph{Management of low-grade glioma}, Curr. Neurol. Neurosci. Rep. \textbf{10}, 224-231 (2010) 

\bibitem{Olson2000} 
Olson, J.D., Riedel E., \& DeAngelis,  L. M. (2000) Long-term outcome of low-grade oligodendroglioma and mixed glioma. \emph{Neurology} \textbf{54}, 1442-1448.

\bibitem{Batchelor2006}
Grier, J. T. \& Batchelor, T. (2006) Low-Grade Gliomas in Adults. \emph{The Oncologist}, \textbf{11}, 681-693.

\bibitem{Smith2008} \textsc{Smith, J.S., Chang, E.F., Lamborn, K.R., Chang, S.M., Prados, M.D., Cha, S., Tihan, T., Vandenberg, S., McDermott, M.W. \& Berger, M.S.} \emph{ Role of extent of resection in the long-term outcome of low-grade hemispheric gliomas.},  \emph{J. Clin. Oncol.} \textbf{26}, 1338-1345 (2008).

\bibitem{Jakola2012}
Jakola, A.S., Myrmel, K.S., Kloster, R., Torp, S.H., Lindal, S., Unsgard, G., Solheim, O.,
\emph{Comparison of a strategy favoring early surgical resection vs a strategy favoring watchful waiting in low-grade gliomas}, 
JAMA \textbf{308}, 1881-1888 (2012).

\bibitem{SS1} J. D. Murray, \emph{Mathematical Biology I and II}, 3rd ed,  Springer (2003).

\bibitem{SS2} E. Mandonnet, J-Y Delattre, M-L Tanguy, K. R. Swanson, A. F. Carpentier, H. Duffau, P. Cornu, R. Van Effenterre, E. C. Alvord Jr., and L. Capelle, \emph{Continuous growth of mean tumor diameter in a subset of WHO grade II gliomas}, Annals of Neurology, \textbf{53}, 524-528 (2003).

\bibitem{SS3} S. Jbabdi, E. Mandonnet, H. Duffau, L. Capelle, K. R. Swanson, M. Pelegrini-Issac, R. Guillevin, H. Benali, \emph{Simulation of anisotropic growth of low-grade gliomas using diffusion tensor imaging},  Magnetic Resonance in Medicine, \textbf{54}, 616-624 (2005).

 \bibitem{Badoual} C. Gerin, J. Pallud, B. Grammaticos, E. Mandonnet, C. Deroulers, P. Varlet, L. Capelle, L. Taillandier, L. Bauchet, H. Duffau, M. Badoual,
\emph{Improving the time-machine: estimating date of birth of grade II gliomas.} Cell. Prolif. \textbf{45}, 76-90 (2012).

 \bibitem{SWCR2} K. R. Swanson, R. C. Rockne, J. Claridge, M. A. Chaplain, E. C. Alvord Jr, A. R. Anderson, \emph{Quantifying the role of angiogenesis in malignant progression of gliomas: in silico modeling integrates imaging and histology.}, Cancer Research \textbf{71}, 7366-7375 (2011).

 \bibitem{PGRT} V. M. P\'erez-Garc\'{\i}a, M. Bogdanska, A. Mart\'{\i}nez-Gonz\'alez, J. Belmonte, P. Schucht, L. A. P\'erez-Romasanta, \emph{Delay effects in the response of low grade gliomas to radiotherapy: A mathematical model and its therapeutical implications}, Mathematical Medicine and Biology (submitted) (2013). 
 
\bibitem{trial6}
Garcia, D.M., Fulling, K.H. \& Marks, J.E., \emph{The value of radiation therapy in addition to surgery for astrocytomas of the adult cerebrum}, Cancer, \textbf{55}, 919-927 (1985).

\bibitem{VandenBent2005}
Van den Bent, M.J., Afra, D., de Witte, O., Ben Hassel, M., Schraub, S., Hoang-Xuan, K., Malmstr\"om, P.O., Collette, L., Pi\'erart, M., Mirimanoff, R. \& Karim, A.B., \emph{Long-term efficacy of early versus delayed radiotherapy for low-grade astrocytoma and oligodendroglioma in adults: the EORTC 22845 randomised trial.},  
Lancet \textbf{366}, 985-990 (2005).

 \bibitem{trial5}
Higuchi Y., Iwadate Y. \& Yamaura A.,  \emph{Treatment of low-grade oligodendroglial tumors without radiotherapy.}, Neurology \textbf{63}, 2384-2386 (2004).

\bibitem{Karim} A. B. M. Karim, B. Maat, et al., \emph{A randomized trial on dose-response in radiation therapy of low-grade cerebral glioma: EORTC study 22844}, Int. J. Rad. Oncol. Biol. Phys. \textbf{36}, 549-556 (1996).

\bibitem{Powatil2007} G Powathil, M Kohandel, S Sivaloganathan, A Oza and M Milosevic, \emph{Mathematical modeling of brain tumors: effects of radiotherapy and chemotherapy}, Phys. Med. Biol. \textbf{52},  3291-3306  (2007)

\bibitem{Rockne2010}
Rockne, R., Hendrickson, K.,  Lai, A.,  Cloughesy, T.,  Alvord, E.C. Jr \& Swanson, K.R. \emph{Predicting the efficacy of radiotherapy in individual glioblastoma patients in vivo: a mathematical modeling approach}, Phys. Med. Biol.  \textbf{55},  3271-3285  (2010).

\bibitem{BondiauRT} %
 Bondiau, P.Y., Frenay, M. \& Ayache, N., \emph{Biocomputing: numerical simulation of glioblastoma growth using diffusion tensor imaging}, Phys. Med. Biol. \textbf{53}, 879-893 (2008).

\bibitem{Konokoglu2010}
Konukoglu, E., Clatz, O., Bondiau, P.Y., Delingette, H. \& Ayache, N., \emph{Extrapolating glioma invasion margin in brain magnetic resonance images: suggesting new irradiation margins}, Med. Image Anal. \textbf{14}, 111-125 (2010).

\bibitem{Kirkby2010}
Barazzuol, L., Burnet, N. G., Jena, R., Jones, B., Jefferies, S. J. \& Kirkby, N. F., \emph{A mathematical model of brain tumour response to radiotherapy and chemotherapy considering radiobiological aspects}, Journal of Theoretical Biology, \textbf{262}, 553-565 (2010).

\bibitem{Stamatakos2006} Stamatakos, G. S., Antipas, V. P. \& Uzunoglu, N. K. (2006) Simulating chemotherapeutic schemes in the individualized treatment context: The paradigm of glioblastoma multiforme treated by temozolomide in vivo. {Comp. Biol. Med.} \textbf{36}, 1216-1234.

 \bibitem{Pallud1}  J. Pallud, L. Taillandier, L. Capelle, D. Fontaine, M. Peyre, F. Ducray, H. Duffau, \& E. Mandonnet, \emph{Quantitative Morphological MRI Follow-up of Low-grade Glioma: A Plead for Systematic Measurement of Growth Rates}, Neurosurgery \textbf{71}, 729-740 (2012).

 \bibitem{Pallud2} J. Pallud, J. F. Llitjos, F. Dhermain, P. Varlet, E. Dezamis, B. Devaux, R. Souillard-Scemama, N. Sanai, M. Koziak, P. Page, M. Schlienger, C. Daumas-Duport, J. F. Meder, C. Oppenheim, \& F. X. Roux, \emph{Dynamic imaging response following radiation therapy predicts long-term outcomes for diffuse low-grade gliomas}, Neuro-Oncology, \textbf{14}, 496-505 (2012).
 
\bibitem{Joiner2009} 
Van der Kogel, A., \& Joiner, M., \emph{Basic clinical radiobiology}, Oxford University Press, 2009
 
\bibitem{graywhite} K. R. Swanson, E. C. alvord, J. D. Murray, \emph{Dynamics of a model for brain tumors reveals a small window for therapeutic intervention}, Disc. Cont. Dyn. Sys. B \textbf{4}, 289-295 (2004).

\bibitem{Ayache1}  S. Jbabdi, E. Mandonnet, H. Duffau, L. Capelle, K. Swanson, M. Pelegrini-
Issac, R. Guillevin, H. Benali, \emph{Simulation of anisotropic growth of low-grade
gliomas using diffusion tensor imaging}, Magnetic Reson. in Med. \textbf{54}, 616Ð624 (2005).

\bibitem{Ayache2} E. Konukoglu, O. Clatz, P.Y. Bondiau, H. Delingette, N. Ayache,
\emph{Extrapolating glioma invasion margin in brain magnetic resonance images: suggesting new irradiation margins}, 
Med. Image Anal. \textbf{14}, 111-125 (2010).

\bibitem{Maini} F. S\'anchez-Gardu\~no, P. K. Maini, \emph{Travelling wave phenomena in some degenerate reaction-diffusion equations}, J. Diff. Eqs. {\bf 117}, 281-319 (1995).

\bibitem{JB} J. Belmonte-Beitia, T. E. Wooley, J. G. Scott, P. K. Maini, E. A. Gaffney, \emph{Modelling biological invasions: Individual to population scales at interfaces}, Journal of Theoretical Biology (submitted)

\bibitem{SIAMTW1} P. K. Braznik, J. J Tyson, \emph{On travelling wave solutions of Fisher's equation in two spatial dimensions}, SIAM J. Appl. Math. \textbf{60}, 371-391 (1999).

\end{thebibliography}
\end{document}